# The effects of halide anions on the electroreduction of $CO_2$ to $C_2H_4$: a density functional theory study

Xifei Ma,[a,b] Lu Xing,[a,b] Xiaoqian Yao,[a] Xiangping Zhang,[a] Lei Liu[a,c,d]* and Suojiang Zhang[a]*

[a] Beijing Key Laboratory of Ionic Liquids Clean Process, State Key Laboratory of Multiphase Complex Systems, CAS Key Laboratory of Green Process and Engineering, Institute of Process Engineering, Chinese Academy of Sciences, Beijing 100190, China
[b] University of Chinese Academy of Science, School of Chemical Engineering, Beijing 100049, China
[c] Dalian National Laboratory for Clean Energy, Dalian 116023, China
[d] School of Chemistry and Chemical Engineering, Wuhan Textile University, Wuhan, 430200, China

* Corresponding authors:
Lei Liu, 2021047@wtu.edu.cn; liulei3039@gmail.com;
Suojiang Zhang, sjzhang@ipe.ac.cn;

**Abstract**: The halide anions present in the electrolyte gradually improves the Faradaic efficiencies (FEs) of the multi-hydrocarbon ($C_{2+}$) products for the electrochemical reduction of $CO_2$ over copper (Cu) catalysts in the order of $F^- < Cl^- < Br^- < I^-$. However, the mechanism behind the increased yield of $C_{2+}$ products with the addition of halide anions still remains indistinct. In this study, we analysed the mechanism by investigating the electronic structures and computing the relative free energies of intermediates formed from $CO_2$ to $C_2H_4$ on the Cu (100) facet based on density functional theory (DFT) calculations. The results show that formyl *CHO species from the hydrogenation reaction of the adsorbed *CO acts as the key intermediate, and the C-C coupling reaction occurs preferentially between the *CHO and *CO with the formation of a *CHO-CO intermediate. Subsequently, the free-energy pathway of $C_2H_4$ formation has been proposed, and we found that the presence of halide anions significantly decreases the free energy of the *CHOCH intermediate, and enhances the desorption capacity of $C_2H_4$ in the order of $F^- < Cl^- < Br^- < I^-$. Lastly, the obtained results are rationalized by the Bader charge analysis.

**Keywords**: $CO_2$ reduction; electrochemical catalysis; density functional theory; C-C coupling; halide anions;

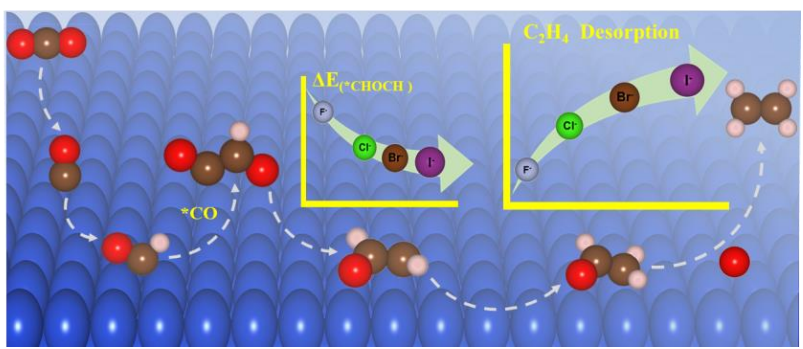

**TOC:** The DFT calculations show that halide anions decrease the free energy of the *CHOCH intermediate and enhance the desorption capacity of $C_2H_4$ from the Cu surface in the order of $F^- < Cl^- < Br^- < I^-$.

## 1. Introduction

The $CO_2$-dominated greenhouse gases arising out of modern society's continue the dependence on fossil fuels, including coal, oil and natural gas, which results in serious climate and environmental issues.[1-2] Therefore, the development of efficient $CO_2$ conversion and utilization technologies is urgently needed. Thermochemistry[3-6], electrochemistry[7-15], photocatalysis[16-18] and bioconversion[19-21] are several attractive strategies for the above-mentioned targets. Among them, the electrochemical reduction of $CO_2$ into value-added chemical products is considered as one of the most promising approaches because of its advantages in terms of mild operating conditions, simplicity of devices and the possibility of being powered by clean electricity, such as the hydro, solar and wind[7-15]. On the other hand, ethylene ($C_2H_4$) is an important organic chemical feedstock used in the synthesis of downstream products like polyethylene (PE), polyvinyl chloride (PVC) and ethylene propylene rubber (EPR). Hence, the production of $C_2H_4$ as the predominant $C_{2+}$ products with high energy density and market value is more profitable compared to $C_1$ products[22-23].

Copper (Cu) is distinctive among metallic materials for its ability to efficiently catalyse the electrochemical reduction of $CO_2$ to multi-hydrocarbon ($C_{2+}$) products. For example, Hori et al. first reported that the faradic efficiency (FE) of $C_2H_4$ was up to 20% with a Cu electrode and in a $KHCO_3$ electrolyte by GC analysis[24], and they found that the FE of $C_2H_4$ was the highest (31.7%) on the (100) facet in comparison with the (110) (15.1%) and (111) (4.7%) facets[25]. Afterwards, several reports also confirmed that $C_2H_4$ formation dominates on the Cu (100) facet[26-28]. Moreover, Hori et al., by using voltammetric and chronopotentiometric approaches, observed that there was a similar product distribution under electrochemical reduction conditions for either $CO_2$ or CO, and concluded that *CO adsorbed on Cu surface was the key intermediate in the electroreduction of $CO_2$[29]. Subsequently, several experimental[30-31] and theoretical[32-33] studies showed that the *CO species act as the key intermediate for the further reduction to $C_2H_4$ over Cu surfaces.

The influence of the electrolyte on the electrochemical reduction of $CO_2$ to $C_2H_4$ is also significant. For example, Hori et al. revealed that the FEs of $C_2H_4$ increased (5.2%, 12.9%, 30.3%, and 30.5%) with increasing ionic radii of alkali metal cations ($Li^+$, $Na^+$, $K^+$, and $Cs^+$) in hydrogencarbonate solutions[34]. Gao et al. reported that the production rate of $C_2H_4$ increased sequentially by adding $Cl^-$, $Br^-$, and $I^-$ to the hydrogencarbonate electrolyte[35], and the authors also found that the geometric partial current density of $C_2H_4$ production (2.6, 16.0, 18.2, 22.5, and 23.6 mA cm$^{-2}$) increased at -1.0 V (vs. RHE) with the variation of anions in the electrolyte ($HCO_3^-$, $CO_3^{2-}$, $Cl^-$, $Br^-$, and $I^-$) based on current-potential plots[36]. Yeo et al. found that the FEs of $C_2H_4$ formation on Cu(100) surface gradually improved (30.6%, 39.5%, 46.9%, and 50.3%) as the anions in the electrolyte varied from $ClO_4^-$, $Cl^-$, $Br^-$ to $I^-$ according to GC measurements[37]. Wang et al. reported that the FEs of $C_2H_4$ formation (27.6%, 30.4%, 31.8%, and 33.2%) were sequentially enhanced by adding the halide anions ($F^-$, $Cl^-$, $Br^-$, and $I^-$) to the KOH electrolyte[23]. As a summary, the experiments demonstrated that the halide anions present in the electrolyte improve the FE of the electrochemical reduction of $CO_2$ to $C_2H_4$ on Cu catalysts in the order of $F^- < Cl^- < Br^- < I^-$.

Presently, it is hypothesized that the specific adsorption of halide anions stabilizes certain intermediates or modifies the surface structure and morphology of the Cu catalyst [35-41]. Nevertheless, the deep mechanism behind the increased yield of $C_2H_4$ with the addition of halide anions still remains indistinct. Based on the density functional theory (DFT) method, we investigated the effects of halide anions on the electroreduction of $CO_2$ to $C_2H_4$, including the stabilities of various possible intermediates, C-C coupling elementary steps, and the entire reaction pathway with halide anions. The DFT calculations show that the main reason is that the halide anion decreases the energy of the *CHOCH intermediate, and facilitates the desorption of $C_2H_4$ from the Cu surface in the order of $F^- < Cl^- < Br^- < I^-$. Subsequent Bader charge analysis indicate that the halide anions change the performance of the Cu catalyst through electrons transfer with the Cu surface.

## 2. Methodology

All first-principle calculations were carried out with the Vienna Ab-initio Simulation Package[42] (VASP) using the plane-wave density functional theory (DFT). The generalized-gradient-approximation (GGA) parameterized by Perdew-Burke-Ernzerhof[43] (PBE) exchange-correlation functional based on Projector-augmented wave[44] (PAW) method was adopted. Van der Waals corrections with the D3 method proposed by Grimme et al. together with Becke-Johnson damping (BJ-damping) function was implemented[45]. Among the three low-index crystal orientations for Cu single-crystal including (100), (110) and (111), the Cu (100) facet was chosen in this work because of its high selectivity for ethylene formation[25, 27]. A fcc (100) surface was modeled as the flat surface of Cu. A $p$ (4 x 4) four-layer periodic cell with a lattice constant of 3.64 Å and a vacuum of 15 Å was employed to eliminate the interaction of adsorbed species between adjacent mirror images caused by periodicity. The dipole correction was added. The two top layers of atoms are free to move, while two bottom layers are fixed in their positions to resemble bulk materials. Plane-wave basis sets with 450 eV cut-off energy and the K-points sampling of a 3 x 3 x 1 Γ-centered Monkhorst-Pack grids in the first Brillouin zone. The first-order Methfessel-Paxton method to set the partial occupancies for each orbital together with the width of the smearing in 0.2 eV ensured that the entropy term is less than 1 mV per atom. The convergence criterion of the electronic self-consistent field (SCF) iteration was specified as $1\times10^{-5}$ eV and $1\times10^{-7}$ eV for the geometric configuration optimization and frequencies calculation, respectively, and the ionic relaxation loop stops when the force on each atom is less than 0.02 eV/Å. The ions were updated and moved by the conjugate gradient algorithm. The Bader charge[46] was calculated using the Bader charge Analysis script written by Henkelman et al. We explicitly included halide ions in our calculations with the aim of revealing their effects on the yield of $C_2H_4$. The zero-point energy (ZPE), enthalpy ($\int_0^T C_p \, dT$) and entropy of molecules and adsorbates are obtained after the calculation of vibrational frequencies based on the quasi-harmonic approximation model[47].

**3. Results and Discussion**

Previous studies show that the key step for the reduction of $CO_2$ to $C_2H_4$ is the C-C coupling based on the *CO intermediate[29-33]. Koper et al proposed that there are two possibilities for the *CO to form a C-C bond, which are the direct *CO and *CO coupling, and the coupling between the *CHO and *CHO (the hydrogenation product of *CO)[32-33]. Therefore, we started our DFT calculations from *CO, and considered three pathways of C-C coupling, and the computed relative free energies (ΔG) together with the optimized structures of key intermediates are shown in Figure 1. In the next contents, we will discuss these three pathways individually, including *CO dimerization (pathway **I**), *CO with *CHO coupling (pathway **II**) and *CO with *COH coupling (pathway **III**). According to DFT calculations, the ΔG for dimerization of surface-adsorbed *CO to form *COCO in pathway **I** is 1.09 eV, while the ΔG for further hydrogenation to *CHOCO is -0.30 eV. For pathway **II**, the adsorbed *CO on Cu (100) facet first accepts a hydrogen atom to form the formyl *CHO, which is an endothermic process with a computing ΔG being 0.67 eV. Afterwards, the *CHO coupled with a *CO and formed a *CHOCO, and the computed ΔG for this step is 0.12 eV. For pathway **III**, the adsorbed *CO also accepts a hydrogen atom, but forms a *COH with a higher energy barrier (0.96 eV). Then *COH reacts with a *CO, and forms a *COHCO with needing a free energy of 0.17 eV. As a conclusion, the pathway **II** is energetically preferred with the lowest energy barrier of 0.67 eV, that is, the *CO is firstly hydrogenated to form a *CHO, and further couples with a coming *CO. Note that, the addition of halide anions shows almost no effects on the selection of the pathways. Specifical specking, by adding $F^-$, $Cl^-$, $Br^-$ and $I^-$ in DFT calculations, the computed free energies of *CHO are still the lowest, which are 0.75, 0.68, 0.68, and 0.67 eV, respectively. Moreover, we have also computed the energy barriers for the C-C coupling of *CO-CO and *CHO-CO by using the nudged elastic band (NEB)[48] method, and the results are given in Figure S2. At a level of PBE, the direct *CO-CO coupling shows an energy barrier of 1.16 eV, while *CHO-CO coupling shows an energy barrier of 0.54 eV, indicating that the pathway **II** is also kinetically favorable.

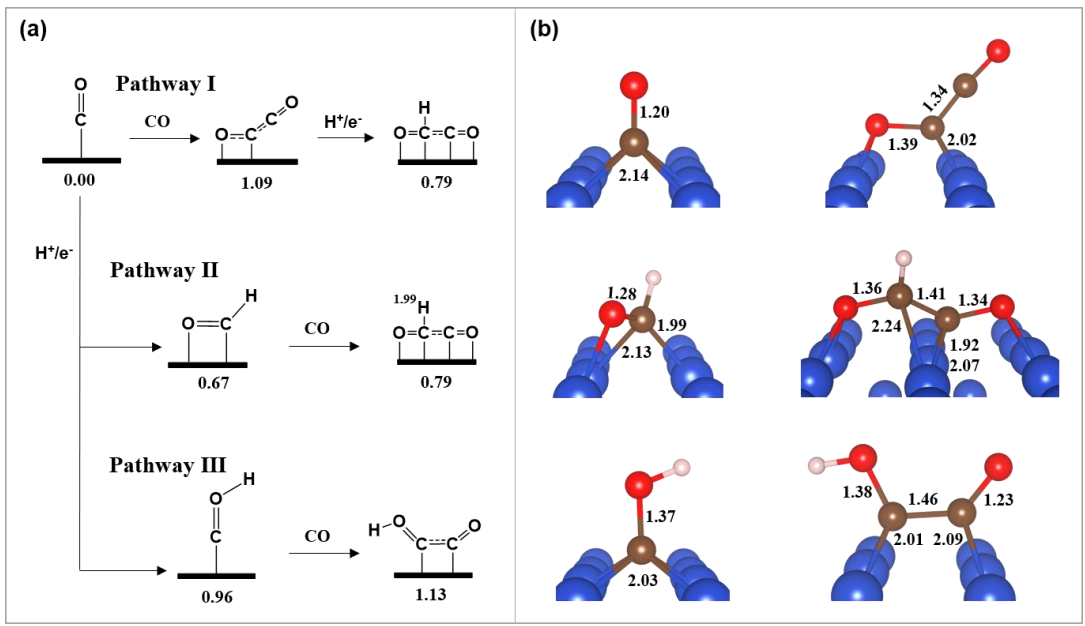

**Figure 1**. (a) Calculated relative free energy (ΔG) of three reaction pathways from the *CO to C-C bond formation on Cu (100) facet without halide ions at $U = 0$ V (vs. RHE) and T = 298.15 K. The energy unit is eV. (b) Optimized structures of key intermediates. The distances are in Å. Color legend: Cu blue, O red, C brown, and H pink.

The optimized structures of key intermediates are depicted in Figure 1(b), and corresponding charge analysis are given in Figure S1. The results show that the adsorbed *CO is bounded to four Cu atoms with the C-Cu distance being 2.14 Å, and the optimized C-O bond is 1.20 Å, which is slightly larger than that in a free CO (1.14 Å). The computed overall valence electrons of *CO is 10.16 $e$. These results suggest that the C-O bond is slightly activated by the electron transfer (ca. 0.16 $e$) through the back-donation from $d$ electrons in metallic Cu to the 2π* orbital of CO[49]. The C-O bonds in *CHO, *COH and *COCO are further weakened, which are 1.28, 1.37 and 1.39 Å, respectively. These values are consistent with the computed free energies for *CHO, *COH and *COCO being 0.67, 0.96 and 1.09 eV, respectively, that is, a geometrically late transition state (TS), often has a higher energy barrier[50]. The computed valance electrons of the *CO, *COH and *CHO are 10.16, 11.38 and 11.40 $e$, respectively. In other words, the *CHO is the most negatively charged intermediates, thus, is the easiest species for CO to attack. Another interested parameter is the C-C bond length along the reaction pathways. The C-C bond length in a free $C_2H_4$ is 1.33 Å (Figure S1). The optimized structure of the dimer *COCO contains a 1.34 Å C-C bond, which is similar to that in $C_2H_4$. However, this species is relatively unstable with a computed free energy of 1.09 eV with respect to the *CO. For *CHOCO and *COHCO coupled products, the optimized C-C distance are similar, which are 1.41 and 1.46 Å, respectively. The difference between those two intermediates is that two C atoms in the former structure are coordinated with two neighbored Cu atom (average C-Cu distance is 2.08 Å), while in the latter one, they are coordinated with two layered Cu atoms (average C-Cu distance is 2.05 Å).

Subsequently, we computed the entire lowest free-energy pathway from $CO_2$ to $C_2H_4$ on Cu (100) facet, following the C-C coupling pathway **II**. The energetical results are summarized in Figure 2, and the detail structural information are given in Figure S3. Here, we started from the adsorbed $CO_2$ (denoted as *$CO_2$), and selected it as the reference state for free energies. Note that the structure of the adsorbed $CO_2$ is identical to the free $CO_2$ molecule in the case of the C-O bond lengths (1.17 Å) and the O-C-O angle (180.0°), and distance between C and Cu atoms is computed to be 3.53 Å. These data indicate that $CO_2$ is mainly physically interacts with the Cu surface, which is similar to that on a Ag surface[51]. After this step, the $CO_2$ is reduced to $C_2H_4$ via transfer of ten electrons and formation of

three H₂O as follows:

1) a free CO$_2$ is adsorbed on Cu (100) surface, and converted to *COOH by adding a hydrogen atom, with the computed ΔG being 0.30 eV. The COOH is bonded to Cu surface through the C atom with a C-Cu distance being 1.94 Å. The bond lengths for C-O and CO(H) are 1.25 and 1.35 Å, respectively, and the angle of O-C-O is 115.6°, which clearly shows that the CO$_2$ is chemically activated. Moreover, we also found some electron transfer from Cu surface to *COOH, of which the total valence electrons are 17.48 *e*; 2) The adsorbed *COOH combines a hydrogen atom to form the surface-bound *CO and release a H$_2$O, which is an exothermic process with a computed ΔG of -0.22 eV. The CO is bonded to the Cu surface through the C atom with four C-Cu distances being 2.14 Å in average, and the bond length of C-O is 1.20 Å. Bader charger analysis shows that there are certain electrons transfer from the Cu surface to *CO, which is 0.16 *e*; 3) The adsorbed *CO accepts a hydrogen atom, and forms the formyl *CHO with a free energy computed to be 0.45 eV. The *CHO is bounded to the Cu surface through the C atom with two C-Cu distances being 1.99 and 2.13 Å, respectively; 4) The adsorbed *CHO couples with *CO, and forms the *CHO-CO, of which the computed free energy is the most positive one along the reaction path (0.56 eV). The *CHOCO is bounded to the Cu surface through the C atom with two C-Cu distances being 1.92 and 2.07 Å, respectively. The bond lengths for C-O and (H)CO are 1.34 and 1.36 Å, respectively, and the bond length for C-C is 1.41 Å. An electron transfer of 0.63 *e* is found from the Cu surface to *CHOCO; 5) The adsorbed *CHOCO is reduced to *CHOCHO (ΔG = -0.13 eV) rather than *CHOHCO (ΔG = 1.02 eV) by adding a hydrogen atom (see details in Figure S3). The CHOCHO is bounded to the Cu surface through the O atom with two O-Cu distances being 1.96 and 2.06 Å, respectively, the bond lengths for the two (H)C-O are 1.34 and 1.35 Å, respectively, and the bond length for C-C is 1.41 Å. A relatively large electron transfer has been found for this elementary step, which is 0.93 *e*; 6) The adsorbed *CHOCHO is further hydrogenated to *CHOCHOH (ΔG = 0.02 eV). The CHOCHOH species is bounded to the Cu surface through the O atom with the O-Cu distance being 2.06 Å. The bond lengths for (H)C-O and (H)C-O(H) are 1.30 and 1.42 Å, respectively, and the bond length for C-C is 1.41 Å; 7) The adsorbed *CHOCHOH is converted to *CHOCH by adding a hydrogen atom and losing a H$_2$O (ΔG = 0.38 eV). The CHOCH species is bounded to the Cu surface through the C atom with two C-Cu distances being 1.95 and 2.05 Å, respectively, and the bond length for C-O(H) is 1.28 Å and for C-C is 1.43 Å. The electron transfer from the Cu surface of this step is computed to be 0.74 $e$; 8) The adsorbed *CHOCH is further hydrogenated to form *OCHCH$_2$ (ΔG = -0.97 eV). The OCHCH$_2$ is bounded to the Cu surface through the O atom with two O-Cu distances being 2.09 and 2.11 Å, respectively, and the bond length for (H)C-O is 1.31 Å and for C-C is 1.40 Å. The *OCHCH$_2$ again attracts some electrons from Cu surface, thus, the computed total valence electrons of this species are 17.63 $e$; 9) The adsorbed *OCHCH$_2$ is reduced to *O_C$_2$H$_4$ by adding a hydrogen atom, which is significantly exothermic (-0.89 eV). The O_C$_2$H$_4$ is bounded to the Cu surface through the C atom with both C-Cu distances being 2.21 Å. The bond length for C-C is 1.38 Å, which is similar to the 1.33 Å of a free C$_2$H$_4$ molecule. Finally, a C$_2$H$_4$ molecule is desorbed from the Cu (100) surface.

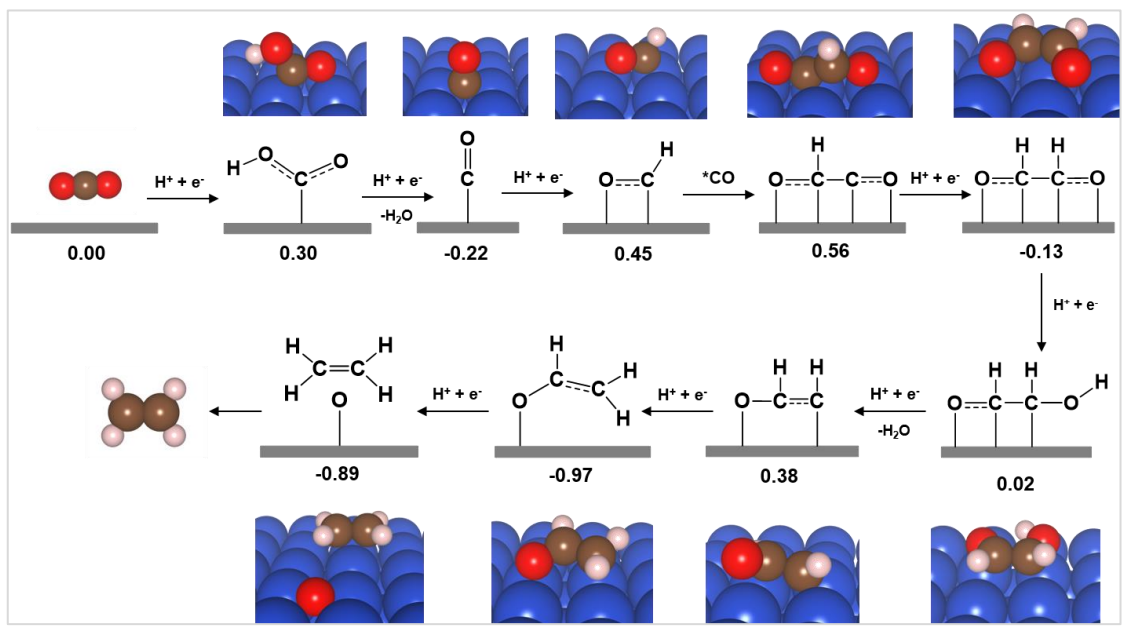

**Figure 2**. Reaction path of the reduction reaction from CO$_2$ to C$_2$H$_4$ on Cu (100) surface at $U = 0$ V (vs. RHE) and T = 298.15 K. The values are the calculated relative free energies (ΔG) with respect to a *CO$_2$, and are given in eV. Color legend: Cu blue, O red, H pink, and C light brown.

Lastly, the halide anions (F$^-$, Cl$^-$, Br$^-$ and I$^-$) are explicitly included in the DFT calculations to study their effects on C$_2$H$_4$ formation following the reaction path shown in Figure 2, and the computed free energies are summarized in Table S1. Here, we will focus on two apparently changed elemental steps, which are the formation of *CHOCH and desorption of $C_2H_4$. The calculated free energies of the *CHOCH on Cu (100) surface in the absence and presence of $F^-$, $Cl^-$, $Br^-$ and $I^-$ anions are 0.38 eV, 0.17 eV, 0.05 eV, 0.04 eV, and 0.05 eV, respectively (Figure 3). These findings indicate that the halide anions present in the electrolyte significantly decreases the formation energy of the *CHOCH, hence, and facilitates its further reduction to the *OCHCH$_2$. Note that the effects from $Cl^-$, $Br^-$ and $I^-$ is more that from $F^-$, which might be one of reasons that the FE of $C_2H_4$ in the case of $F^-$ is the smallest. The computed bonding energy of the *CHOCH with Cu (100) increases in the order of free $X^- < F^- < Cl^- < Br^- < I^-$ being -2.67, -2.87, -2.95, -2.99 and -3.02 eV, respectively (Table S3-4), which is consistent with the order found for the free energy of the *CHOCH. As discussed above, the CHOCH species bound to the empty Cu (100) surface with an average C-Cu distance being 2.27 Å, and the bond length of C-C is 1.43 Å. After the addition of halide anions, we observed several notable geometric changes. For example, the terminal O atom bounds to the empty Cu surface with a distance being 2.18 Å, while no interactions were found in the presence of all halide anions, although we started the geometry optimizations from the almost identical structures. The C-Cu bond length decreases from 2.27 (empty) to be an average bond length of 2.09 Å ($I^-$). Moreover, we found that F coordinates to two Cu atoms with the distance being 2.00 Å, while Cl, Br and I coordinates to four Cu atoms with the distances being 2.50, 2.60, and 2.72 Å, respectively. Subsequently, we performed Bader charge analysis for the *CHOCH without and with halide anions. The results show that the total valence electrons of *CHOCH species adsorbed on Cu (100) in the absence of halide anions and the presence of $F^-$, $Cl^-$, $Br^-$ and $I^-$ are 16.74, 16.82, 16.92, 16.93 and 16.91, respectively. Hence, the electron transfer of 0.74, 0.82, 0.92, 0.93 and 0.91$e$ occurs from Cu surfaces compared to the free CHOCH. While the valence electrons on the F, Cl, Br and I are 7.65, 7.43, 7.27 and 6.93, respectively. Based on those data, we conclude that the presence of halide anions allows more electron transfer between the adsorbed *CHOCH and the Cu surfaces, which changes their interactions and the stabilities of *CHOCH.

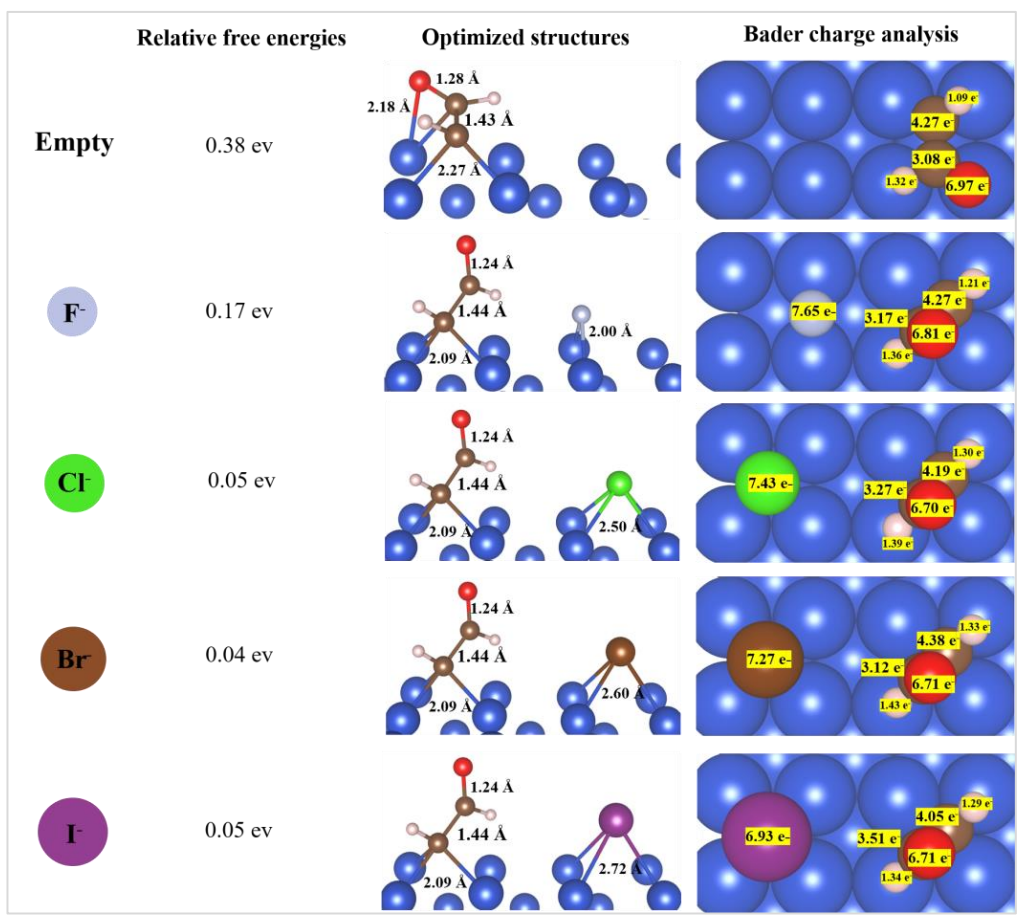

**Figure 3**. Relative free energies, optimized structures and Bader charges of the *CHOCH in the absence and presence of F⁻, Cl⁻, Br⁻ and I⁻. Color legend: Cu blue, F silver, Cl green, Br dark brown, I purple, O red, H pink, and C light brown.

The calculated free energies of *O_$C_2H_4$ on the Cu (100) surface in the absence and presence of F⁻, Cl⁻, Br⁻ and I⁻ anions are -0.89, -0.67, -0.63, -0.56, and -0.43 eV, respectively (Figure 4), indicating that the halide anions present in the electrolyte significantly decrease the interactions between *O_$C_2H_4$ and Cu surfaces, thus, $C_2H_4$ gets easier to be desorbed. This conclusion is also supported by the computed bonding energy of the *$C_2H_4$ with Cu (100), which are -0.81, -0.58, -0.52, -0.51 and -0.48 eV, respectively, for empty, F⁻, Cl⁻, Br⁻ and I⁻ (Table S3-4). The *O_$C_2H_4$ is bounded to one Cu atom with an average C-Cu distance being 2.20 Å, and the bond length of C-C is 1.38 Å. It is found that the addition of halide anions changes the C atom coordination, and the two C atoms are bounded to two different Cu atoms. The C-Cu bond length slightly increases from 2.20 (empty) to an average bond length of 2.25 Å (I⁻). Unlike the case of *CHOCH, we found that all halide anions coordinate to two Cu atoms, with the distances being 2.03, 2.34, 2.49 and 2.65 Å, respectively, for F[-], Cl[-], Br[-] and I[-]. We then performed Bader charge analysis for *O_$C_2H_4$ without and with halide anions. The results show that the adsorbed *$C_2H_4$ attracts 0.20, 0.22, 0.28, 0.34 and 0.38 $e$ from Cu surfaces for empty, F[-], Cl[-], Br[-] and I[-], respectively, making their total valence electrons being 12.20, 12.22, 12.28, 12.34 and 12.38, respectively. The computed valence electrons for F[-], Cl[-], Br[-] and I[-] are 7.63, 7.29, 7.01 and 6.87, respectively. As such, the presence of the halide anions changes the charge population of the Cu surfaces through electron transfer, thus, changes the adsorption capacity of *$C_2H_4$ on the surface.

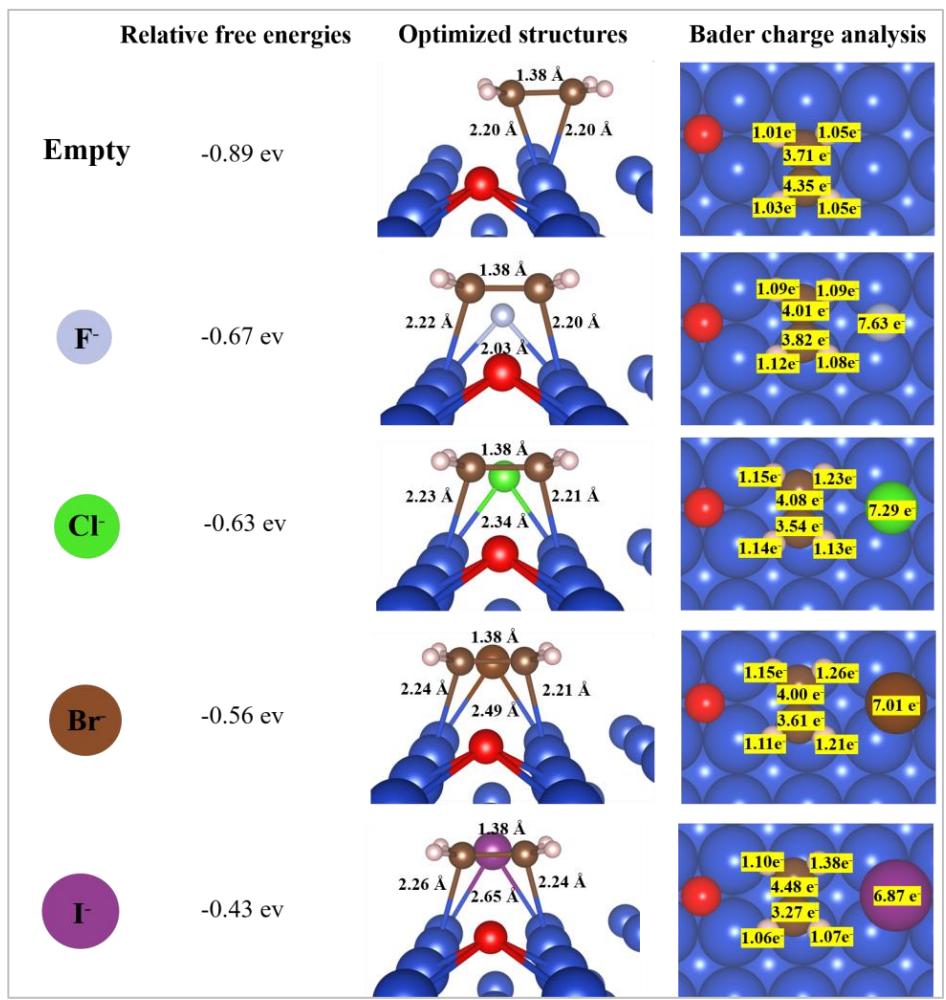

**Figure 4**. Relative free energies, optimized structures and Bader charges of the *O_$C_2H_4$ in the absence and presence of F[-], Cl[-], Br[-] and I[-]. Color legend: Cu blue, F silver, Cl green, Br dark brown, I purple, O red, H pink, and C light brown.

As a summary, the reason behinds improving C$_2$H$_4$ FE in the presence of halide anions is investigated by DFT calculations. On the one hand, the presence of halide anions decreases the free energy of the *CHOCH, and facilitates its further reduction to *OCHCH$_2$. On the other hand, the halide anions are beneficial for the desorption of C$_2$H$_4$ from the Cu surfaces. The Bader charge analysis reveal that the presence of halide anions changes charge populations of the Cu surface. Importantly, those findings are following orders of empty < F$^-$ < Cl$^-$ < Br$^-$ < I$^-$, and they are quite consistent with the experimental observation of the C$_2$H$_4$ FE, which shows an order of empty < F$^-$ < Cl$^-$ < Br$^-$ < I$^-$.

## 4. Conclusions

In this study, DFT calculations were performed to investigate the effects of halide anions on the electroreduction of CO$_2$ to C$_2$H$_4$ on the Cu (100) surface. First, several intermediates and C-C couple pathways were studied by the geometry optimizations as well as free energy calculations. Based on the free energy profile, we found that the formyl *CHO from hydrogenation of the adsorbed *CO is more favourable as the key intermediate for the further reduction, and C-C coupling preferentially proceeds between the adsorbed *CHO and *CO. Secondly, halide anion effects were studied by explicitly including halide anions in the DFT calculations (e.g. F$^-$, Cl$^-$, Br$^-$, and I$^-$). The results show that the presence of halide anions changes the stability of the intermediates and final products on the Cu (100) surface. The addition of halide anions decreases the free energy of *CHOCH, and the interactions of C$_2$H$_4$ in an order of F$^-$ < Cl$^-$ < Br$^-$ < I$^-$. Lastly, depth analysis into halide anion effects were performed by the charge analysis, and it is shown that adsorption of halide anions changes the charge populations of Cu surfaces through electron transfer. From a solvent point of view, the presented DFT calculations reveals possible reasons for the effects of halide anions on the electrochemical reduction of CO$_2$ to C$_2$H$_4$ on the Cu catalysts. As a perspective, the reactions between halide anions with Cu surface should be also considered.

## Supporting Information

Activation energy barrier calculations for *CO-CO and *CHO-CO couplings, relative free energies of elementary reactions from CO$_2$ to C$_2$H$_4$, the optimized structures of intermediates along the lowest free-energy pathway.

**Conflicts of interest**

There are no conflicts to declare.

**Acknowledgments**

This work was financially supported by the DNL Cooperation Fund, CAS (DNL202007), and National Natural Science Foundation of China (No. 21776281).

**References**

1. Kintisch, E., Climate Crossroads. *Science* **2015**, *350*, 1016-1017.
2. Li, B. W.; Wang, C. L.; Zhang, Y. Q.; Wang, Y. L., High $CO_2$ Absorption Capacity of Metal-Based Ionic Liquids: A Molecular Dynamics Study. *Green Energy Environ.* **2021**, *6*, 253-260.
3. Heenemann, M.; Millet, M. M.; Girgsdies, F.; Eichelbaum, M.; Risse, T.; Schlogl, R.; Jones, T.; Frei, E., The Mechanism of Interfacial $CO_2$ Activation on Al Doped Cu/Zno. *ACS Catal.* **2020**, *10*, 5672-5680.
4. Wang, S. W.; Wu, T. J.; Lin, J.; Ji, Y. S.; Yan, S. R.; Pei, Y. T.; Xie, S. H.; Zong, B. N.; Qiao, M. H., Iron-Potassium on Single-Walled Carbon Nanotubes as Efficient Catalyst for $CO_2$ Hydrogenation to Heavy Olefins. *ACS Catal.* **2020**, *10*, 6389-6401.
5. Xu, D.; Ding, M. Y.; Hong, X. L.; Liu, G. L.; Tsang, S. C. E., Selective C-2(+) Alcohol Synthesis from Direct $CO_2$ Hydrogenation over a Cs-Promoted Cu-Fe-Zn Catalyst. *ACS Catal.* **2020**, *10*, 5250-5260.
6. Sun, W. Z.; Wang, M. C.; Zhang, Y. Q.; Ding, W. L.; Huo, F.; Wei, L.; He, H. Y., Protic Vs Aprotic Ionic Liquid for $CO_2$ Fixation: A Simulation Study. *Green Energy Environ.* **2020**, *5*, 183-194.
7. Albo, J.; Alvarez-Guerra, M.; Castano, P.; Irabien, A., Towards the Electrochemical Conversion of Carbon Dioxide into Methanol. *Green Chem.* **2015**, *17*, 2304-2324.
8. Ebaid, M.; Jiang, K.; Zhang, Z. M.; Drisdell, W. S.; Bell, A. T.; Cooper, J. K., Production of C-2/C-3 Oxygenates from Planar Copper Nitride-Derived Mesoporous Copper Via Electrochemical Reduction of $CO_2$. *Chem. Mater.* **2020**, *32*, 3304-3311.
9. Gao, D. F.; Zhou, H.; Wang, J.; Miao, S.; Yang, F.; Wang, G. X.; Wang, J. G.; Bao, X. H., Size-Dependent Electrocatalytic Reduction of $CO_2$ over Pd Nanoparticles. *J. Am. Chem. Soc.* **2015**, *137*, 4288-4291.
10. Gao, S.; Lin, Y.; Jiao, X. C.; Sun, Y. F.; Luo, Q. Q.; Zhang, W. H.; Li, D. Q.; Yang, J. L.; Xie, Y., Partially Oxidized Atomic Cobalt Layers for Carbon Dioxide Electroreduction to Liquid Fuel. *Nature* **2016**, *529*, 68-+.
11. Han, N.; Ding, P.; He, L.; Li, Y. Y.; Li, Y. G., Promises of Main Group Metal-Based Nanostructured Materials for Electrochemical $CO_2$ Reduction to Formate. *Adv. Energy Mater.* **2020**, *10*.
12. Ma, L. S., et al., Covalent Triazine Framework Confined Copper Catalysts for Selective Electrochemical Co2 Reduction: Operando Diagnosis of Active Sites. *ACS Catal.* **2020**, *10*, 4534-4542.
13. Qiao, J. L.; Liu, Y. Y.; Hong, F.; Zhang, J. J., A Review of Catalysts for the Electroreduction of Carbon Dioxide to Produce Low-Carbon Fuels. *Chem. Soc. Rev.* **2014**, *43*, 631-675.

14. Whipple, D. T.; Kenis, P. J. A., Prospects of $CO_2$ Utilization Via Direct Heterogeneous Electrochemical Reduction. *J. Phys. Chem. Lett.* **2010**, *1*, 3451-3458.

15. Wu, J. J.; Sharifi, T.; Gao, Y.; Zhang, T. Y.; Ajayan, P. M., Emerging Carbon-Based Heterogeneous Catalysts for Electrochemical Reduction of Carbon Dioxide into Value-Added Chemicals. *Adv. Mater.* **2019**, *31*.

16. Hao, L.; Kang, L.; Huang, H. W.; Ye, L. Q.; Han, K. L.; Yang, S. Q.; Yu, H. J.; Batmunkh, M.; Zhang, Y. H.; Ma, T. Y., Surface-Halogenation-Induced Atomic-Site Activation and Local Charge Separation for Superb $CO_2$ Photoreduction. *Adv. Mater.* **2019**, *31*.

17. Kumaravel, V.; Bartlett, J.; Pillai, S. C., Photoelectrochemical Conversion of Carbon Dioxide ($CO_2$) into Fuels and Value-Added Products. *ACS Energy Lett.* **2020**, *5*, 486-519.

18. Li, Y.; Li, B. H.; Zhang, D. N.; Cheng, L.; Xiang, Q. J., Crystalline Carbon Nitride Supported Copper Single Atoms for Photocatalytic $CO_2$ Reduction with Nearly 100% CO Selectivity. *ACS Nano* **2020**, *14*, 10552-10561.

19. Kumar, A.; Ergas, S.; Yuan, X.; Sahu, A.; Zhang, Q. O.; Dewulf, J.; Malcata, F. X.; van Langenhove, H., Enhanced $CO_2$ Fixation and Biofuel Production Via Microalgae: Recent Developments and Future Directions. *Trends Biotechnol.* **2010**, *28*, 371-380.

20. Lam, M. K.; Lee, K. T., Microalgae Biofuels: A Critical Review of Issues, Problems and the Way Forward. *Biotechnol. Adv.* **2012**, *30*, 673-690.

21. Zeng, X. H.; Danquah, M. K.; Chen, X. D.; Lu, Y. H., Microalgae Bioengineering: From $CO_2$ Fixation to Biofuel Production. *Renewable Sustainable Energy Rev.* **2011**, *15*, 3252-3260.

22. Dinh, C. T., et al., $CO_2$ Electroreduction to Ethylene Via Hydroxide-Mediated Copper Catalysis at an Abrupt Interface. *Science* **2018**, *360*, 783-787.

23. Ma, W. C.; Xie, S. J.; Liu, T. T.; Fan, Q. Y.; Ye, J. Y.; Sun, F. F.; Jiang, Z.; Zhang, Q. H.; Cheng, J.; Wang, Y., Electrocatalytic Reduction of $CO_2$ to Ethylene and Ethanol through Hydrogen-Assisted C-C Coupling over Fluorine-Modified Copper. *Nat. Catal.* **2020**, *3*, 478-487.

24. Hori, Y.; Kikuchi, K.; Murata, A.; Suzuki, S., Production of Methane and Ethylene in Electrochemical Reduction of Carbon Dioxide at Copper Electrode in Aqueous Hydrogencarbonate Solution. *Chemistry Letters* **1986**, 897-898.

25. Hori, Y.; Wakebe, H.; Tsukamoto, T.; Koga, O., Adsorption of CO Accompanied with Simultaneous Charge-Transfer on Copper Single-Crystal Electrodes Related with Electrochemical Reduction of $CO_2$ to Hydrocarbons. *Surface Science* **1995**, *335*, 258-263.

26. Jiang, K.; Sandberg, R. B.; Akey, A. J.; Liu, X. Y.; Bell, D. C.; Norskov, J. K.; Chan, K. R.; Wang, H. T., Metal Ion Cycling of Cu Foil for Selective C-C Coupling in Electrochemical $CO_2$ Reduction. *Nat. Catal.* **2018**, *1*, 111-119.

27. Hori, Y.; Takahashi, I.; Koga, O.; Hoshi, N., Selective Formation of C2 Compounds from Electrochemical Reduction of $CO_2$ at a Series of Copper Single Crystal Electrodes. *J. Phys. Chem. B* **2002**, *106*, 15-17.

28. Hori, Y.; Takahashi, I.; Koga, O.; Hoshi, N., Electrochemical Reduction of Carbon Dioxide at Various Series of Copper Single Crystal Electrodes. *Journal of Molecular Catalysis a-Chemical* **2003**, *199*, 39-47.

29. Hori, Y.; Murata, A.; Yoshinami, Y., Adsorption of CO, Immediately Formed in Electrochemical Reduction of $CO_2$, at a Copper Electrode. *Journal of the Chemical Society-Faraday Transactions* **1991**, *87*,

125-128.

30. Hori, Y.; Takahashi, R.; Yoshinami, Y.; Murata, A., Electrochemical Reduction of CO at a Copper Electrode. *J. Phys. Chem. B* **1997**, *101*, 7075-7081.

31. Wang, L., et al., Electrochemical Carbon Monoxide Reduction on Polycrystalline Copper: Effects of Potential, Pressure, and Ph on Selectivity toward Multicarbon and Oxygenated Products. *ACS Catal.* **2018**, *8*, 7445-7454.

32. Schouten, K. J. P.; Kwon, Y.; van der Ham, C. J. M.; Qin, Z.; Koper, M. T. M., A New Mechanism for the Selectivity to C-1 and C-2 Species in the Electrochemical Reduction of Carbon Dioxide on Copper Electrodes. *Chem. Sci.* **2011**, *2*, 1902-1909.

33. Schouten, K. J. P.; Qin, Z. S.; Gallent, E. P.; Koper, M. T. M., Two Pathways for the Formation of Ethylene in CO Reduction on Single-Crystal Copper Electrodes. *J. Am. Chem. Soc.* **2012**, *134*, 9864-9867.

34. Murata, A.; Hori, Y., Product Selectivity Affected by Cationic Species in Electrochemical Reduction of $CO_2$ and CO at Cu Electrode. *Bull. Chem. Soc. Jpn.* **1991**, *64*, 123-127.

35. Gao, D. F.; Sohoken, F.; Roldan Cuenya, B., Improved $CO_2$ Electroreduction Performance on Plasma-Activated Cu Catalysts Via Electrolyte Design: Halide Effect. *ACS Catal.* **2017**, *7*, 5112-5120.

36. Gao, D. F.; Sinev, I.; Scholten, F.; Aran-Ais, R. M.; Divins, N. J.; Kvashnina, K.; Timoshenko, J.; Roldan Cuenya, B., Selective $CO_2$ Electroreduction to Ethylene and Multicarbon Alcohols Via Electrolyte-Driven Nanostructuring. *Angew. Chem., Int. Ed.* **2019**, *58*, 17047-17053.

37. Huang, Y.; Ong, C. W.; Yeo, B. S., Effects of Electrolyte Anions on the Reduction of Carbon Dioxide to Ethylene and Ethanol on Copper (100) and (111) Surfaces. *ChemSusChem* **2018**, *11*, 3299-3306.

38. Varela, A. S.; Ju, W.; Reier, T.; Strasser, P., Tuning the Catalytic Activity and Selectivity of Cu for $CO_2$ Electroreduction in the Presence of Halides. *ACS Catal.* **2016**, *6*, 2136-2144.

39. Yano, H.; Tanaka, T.; Nakayama, M.; Ogura, K., Selective Electrochemical Reduction of $CO_2$ to Ethylene at a Three-Phase Interface on Copper(I) Halide-Confined Cu-Mesh Electrodes in Acidic Solutions of Potassium Halides. *J. Electroanal. Chem.* **2004**, *565*, 287-293.

40. Lee, W. H.; Byun, J.; Cho, S. K.; Kim, J. J., Effect of Halides on Cu Electrodeposit Film: Potential-Dependent Impurity Incorporation. *J. Electrochem. Soc.* **2017**, *164*, D493-D497.

41. Gao, D. F.; McCrum, I. T.; Deo, S.; Choi, Y. W.; Scholten, F.; Wan, W. M.; Chen, J. G. G.; Janik, M. J.; Roldan Cuenya, B., Activity and Selectivity Control in $CO_2$ Electroreduction to Multicarbon Products over Cuox Catalysts Via Electrolyte Design. *ACS Catal.* **2018**, *8*, 10012-+.

42. Kresse, G.; Furthmuller, J., Efficient Iterative Schemes for Ab Initio Total-Energy Calculations Using a Plane-Wave Basis Set. *Phys. Rev. B* **1996**, *54*, 11169-11186.

43. Perdew, J. P.; Burke, K.; Ernzerhof, M., Generalized Gradient Approximation Made Simple. *Phys. Rev. Lett.* **1996**, *77*, 3865-3868.

44. Kresse, G.; Joubert, D., From Ultrasoft Pseudopotentials to the Projector Augmented-Wave Method. *Phys. Rev. B* **1999**, *59*, 1758-1775.

45. Grimme, S.; Ehrlich, S.; Goerigk, L., Effect of the Damping Function in Dispersion Corrected Density Functional Theory. *J. Comput. Chem.* **2011**, *32*, 1456-1465.

46. Tang, W.; Sanville, E.; Henkelman, G., A Grid-Based Bader Analysis Algorithm without Lattice Bias. *Journal of Physics-Condensed Matter* **2009**, *21*.

47. Bahn, S. R.; Jacobsen, K. W., An Object-Oriented Scripting Interface to a Legacy Electronic Structure

Code. *Comput. Sci. Eng.* **2002**, *4*, 56-66.

48. Henkelman, G.; Jonsson, H., Improved Tangent Estimate in the Nudged Elastic Band Method for Finding Minimum Energy Paths and Saddle Points. *J. Chem. Phys.* **2000**, *113*, 9978-9985.

49. Blyholder, G., Molecular Orbital View of Chemisorbed Carbon Monoxide. *J. Phys. Chem.* **1964**, *68*, 2772-&.

50. Liu, L.; Vankova, N.; Heine, T., A Kinetic Study on the Reduction of $CO_2$ by Frustrated Lewis Pairs: From Understanding to Rational Design. *Phys. Chem. Chem. Phys.* **2016**, *18*, 3567-3574.

51. Guo, S.; Li, Y.; Liu, L.; Liu, L. C.; Zhang, X. P.; Zhang, S. J., Computational Identification of a New Adsorption Site of CO(2)on the Ag (211) Surface. *Chemistryselect* **2020**, *5*, 11503-11509.